# Investigation of the Quantum Vacuum as an Energy Sink for Subcritical and Supercritical Vaporization Lasers

Jeffrey S. Lee[1]
Gerald B. Cleaver[1,2]

[1]Early Universe Cosmology and Strings Group
Center for Astrophysics, Space Physics, and Engineering Research
[2]Department of Physics
Baylor University
One Bear Place
Waco, TX 76706

Jeff_Lee@Baylor.edu
Gerald_Cleaver@Baylor.edu

## 1. Abstract

In this paper, it is shown that the quantum electrodynamic vacuum particle production rate by a vaporization laser is negligible and is not a significant energy sink at electric field strengths beyond the Schwinger Limit.

## 2. Introduction

Recent advances in laser technology have resulted in intensities sufficient for vacuum polarization. Even though energy extraction by the QED vacuum is considered to be in a one-to-one energy balance with absorbed energy, the potential concern that the quantum vacuum could serve as a significant energy sink for vaporization lasers with intensities beyond the Schwinger Limit is worth addressing.

Considered in this paper is the case of a subcritical (below the Schwinger Limit) laser which generates a particle production rate that represents only an extremely negligible loss of energy from the system. Also examined are hypothetical 1 μm ($10^{-6}$ m) and 1 fm ($10^{-15}$ m) lasers capable of pulse intensities significantly in excess of the Schwinger Limit. In all cases, however, the power loss due to bleeding the quantum electrodynamic vacuum is shown to be extremely negligible.

## 3. One-to-One Energy: A Hawking Radiation Analogy

Since the total energy emitted by a black hole as Hawking radiation throughout its lifetime never exceeds the initial mass energy of the black hole and the sum of all ingested energies, there is no apparent physical justification for a laser with an **E**-field in excess of the Schwinger Limit $\left(E_S = \frac{m_e^2 c^3}{e\hbar} \approx 1.3 \times 10^{18} \text{ V/m}\right)$[1] being capable of releasing more energy through particle production than it delivers. A release of greater energy from the QED vacuum than was incident from the laser would initiate a continually increasing energy feedback loop, and the QED vacuum would propagate a radially expanding luminal disturbance from the laser spot. Thus, if sufficient energy density, either photonic or gravitational, exists in a region of spacetime, then particles will be produced accordingly, but never with a total energy in excess of the relinquished energy. The stored potential energy being released with a small activation energy, as in the case of a conventional or nuclear explosive, is not applicable to particle production by a laser (or a black hole), and the analogy between particle production and the release of an explosive's potential energy would not be valid.

## 4. Particle Production by Sub-Critical Fields

Vacuum pair production is an elegantly simple method of testing the validity of QED, particularly in the subcritical field regime. High energy photon-photon scattering is an instrumental process in determining the optical opacity of the interstellar and intergalactic media [1]. Pair production provides a mechanism not only by which positrons are generated in the purlieu of active galactic nuclei, but also by which primordial black holes generate multiple particle species (Hawking radiation) in the region surrounding their event horizons.

Gregori et al. point out that "In simpler terms, as particles are created, their associated electric field adds to the external field, which then feeds back to the production of the next pair" [1]. Although this speculation lacks detailed theoretical studies, in classical electromagnetic theory the direction of the electric field would be from the positron to the electron. However, the direction of motion of the electron is opposite to the direction of the external field, while that of the positron is in the same direction as the external field (i.e. the direction from the $e^+$ to the $e^-$ is antiparallel to the direction of the external field). Consequently, the "associated electric field" direction is opposite to the external electric field's direction, and hence, the "feed-back strengthen" mechanism is actually a "feed-back weaken" mechanism. This effect will have the consequence of depleting the particle production rate, and the energy loss from the laser will be consequently reduced.

Although they currently lack experimental verification, quantum mean field approaches by which vacuum breakdown occurs as the result of numerous low energy photons, have been suggested [2], [3], [4]. In spinor QED, the Quantum Vlasov Equation (QVE) (eq. (1)) gives the particle number operator for fermions in the presence of a semi-classical electric field.

$$\frac{df_k(t)}{dt} = \frac{\dot{\Omega}_k(t)\varepsilon_\perp(t)}{2\Omega_k(t)\varepsilon_\parallel(t)} \times \int_{-\infty}^{t} d\tau \left\{ \frac{\dot{\Omega}_k(t)\varepsilon_\perp(t)}{2\Omega_k(t)\varepsilon_\parallel(t)} [1 - 2f_k(u)] cos\left[2\int_u^t d\tau \Omega_k(\tau)\right] \right\} \quad (1)$$

---

[1] $E_S$ is the Schwinger Limit electric field; $m_e$ is the electron/positron mass; $c$ is the speed of light; $e$ is the fundamental charge; $\hbar$ is the Reduced Planck's Constant.

where

$$\Omega_k^2 = m^2 + \varepsilon_\perp^2 + (k_\parallel - eA)^2 \qquad (2)$$

and

$$\varepsilon_\perp^2 = m^2 + k_\perp^2, \qquad (3)$$

$$\varepsilon_\parallel^2 = (k_\parallel - eA)^2. \qquad (4)$$

The momenta which are perpendicular and parallel to the linearly polarized electric field $\left(\varepsilon = -\dot{A}\right)$ in the Coulomb gauge are $\varepsilon_\perp$ and $\varepsilon_\parallel$ respectively. The validity that this work is within the Coulomb gauge is supported by the use of Kim's and Page's approach, in which they make use of the phase-integral method [5], [6] to develop further the instanton method of refs. [7], [8]. By defining the instanton action in a gauge-independent way as a contour integral in the complex space or time plane, the choice of the Coulomb gauge remains justifiable.

Integrating over all momenta yields the total electron-positron volumetric number density (eq. (5)),

$$N(t) = 2 \int \frac{d^3 k}{(2\pi)^3} f_k(t). \qquad (5)$$

It is noteworthy that the $[1 - 2 f_k(u)]$ factor in eq. (1) gives rise to the non-Markovian nature of the QVE because particles which are already present in the system, and initially entangled, will affect the pair production rate. Although the numerous approaches to subcritical particle production [9], [10], [11], [12], [13], [14], [15], [16], including the Non-equilibrium Quantum Field Theory (NeqQFT) framework, produce slightly different results, these approaches appear to converge for electric fields in excess of the Schwinger Limit [17]. Additionally, the optimistic case of particle production, given by the QVE, is simultaneously the worst-case scenario for laser energy loss. It is for this reason that the QVE is being considered.

It is noteworthy that this theory considers the QVE as an ordinary differential equation of the probability density function $f$, but it lacks the essential internal mathematical structure requiring the constraint that $f \geq 0$ for all $k$ and $t$. Specifically, the time-evolution of $f$, from a non-negative initial function, could yield a negative-value regime if the QVE determines the specific probability density function (i.e. it would be necessary to determine that $f \geq 0$ across the interval $t = -\infty$ to $t = \infty$).[2]

Additionally, for intensities in excess of $10^{21}$ W/cm$^2$, the laser pulse interaction with any ultra-relativistic electrons and positrons that are produced will result in the production of gamma ray photons and additional particle pairs [18]. However, the "feed-back weaken" mechanism (discussed above) is expected to prevent this process from appreciably enhancing particle production and significantly depleting the laser energy.

For static fields, Schwinger originally showed an exponential suppression of the pair number, which in the case of a temporally static field, is given by eq. (6).

---

[2] Although cautious readers may be uncomfortable about trusting the QVE, and would prefer the Klein-Gordon Equation (in which the probability density function is required to have non-zero squared-values), the rates of particle production, and hence the laser energy depletion rates, of the two methodologies are expected to be comparable.

$$N_{\text{Schwinger}} = \exp\left(-\pi \frac{\varepsilon_S}{\varepsilon}\right) \tag{6}$$

where $\varepsilon_S$ is the critical (Schwinger) field ($1.3 \times 10^{18}$ V/m). However, even for EM fields approaching and exceeding the Schwinger Limit, it is not expected that the electron-positron pair production rate by the Breit-Wheeler process [19] will be enhanced significantly.

Simulation results from the Astra Gemini and the Vulcan PW systems located at the Rutherford Appleton Laboratory establish that the expected γ-γ pair production rates represent only very negligible energy losses, as shown in Table 1.

|  | **Astra Gemini** | **Vulcan PW** |
|---|---|---|
| Wavelength (nm) | 800 | 1064 |
| Pulse Length (fs) | 30 | 500 |
| Laser Energy (J) | 15 | 500 |
| Spot Diameter (μm) | 5 | 5 |
| Intensity (W/cm$^2$) | $2.5 \times 10^{21}$ | $5 \times 10^{21}$ |
| $\varepsilon_o$ (V/m) | $1.4 \times 10^{14}$ | $1.9 \times 10^{14}$ |
| $n_{av}$ (cm$^3$) | $8 \times 10^{20}$ | $1.6 \times 10^{21}$ |
| $N_{ep}$ | $1.6 \times 10^{10}$ | $4.2 \times 10^{10}$ |
| $\tau_{mi}$ (fs) | $9.9 \times 10^{-10}$ | $5.1 \times 10^{-10}$ |
| $\tau_{qu}$ (fs) | $8.1 \times 10^{-6}$ | $8.1 \times 10^{-6}$ |
| $\tau_{cl}$ (fs) | $1.2 \times 10^{-2}$ | $8.7 \times 10^{-3}$ |
| $\Delta N_{ep}$ | $2.6 \times 10^3$ | $5.0 \times 10^3$ |
| $2\pi\tau_{cl}/v$ | 0.13 | 0.22 |
| $N^{\gamma\gamma}$ | 0.63 | 0.21 |
| Repetition Rate | every 20 seconds | every 1 hour |
| $N^{\gamma\gamma}$ after 10 hr | 10879 | 805 |

Table 1: Operation parameters for the Rutherford Appleton Laboratory's Astra Gemini and the Vulcan PW laser systems and expected γ-γ yield. A beam crossing angle of $\theta = 135°$ has been assumed. **[1]**

The worldwide increase in (peak power) Petawatt lasers has been significant in recent years with more than five 10 PW lasers currently operating [20], [21], [22], and the research applications are significant [23]. Among these lasers are The Scalable High average-power Advanced Radiographic Capability (SHARC), the Big Aperture Thulium (BAT), and the L4-ATON which is designed to deliver 10 PW laser pulses in excess of 1.5 kJ with pulse times less than 150 fs and at a repetition rate up to 1 shot/minute [24]. SHARC boasts an average power of 1.5 kW for the creation of *x*-rays and ion beams. The above-mentioned lasers, as well as the Apollon 10 PW laser with intensities surpassing $2 \times 10^{22}$ W/cm$^2$ (150 J of laser energy, a 15 fs pulse duration, and a repetition rate of 1 shot/min) [25] generate only an inconsequential increase in $N^{\gamma\gamma}$.

## 5. Particle Production from a Vaporization Laser

Significant advances in chirped pulse amplification and *x*-ray free electron and gamma ray lasers have made possible intense subcritical electric fields which exceed the nonlinear Compton scattering regime and enter the vacuum birefringence regime. Lasers capable of Schwinger Limit intensities could be realized by the year 2030 (see Figure 1) [26].

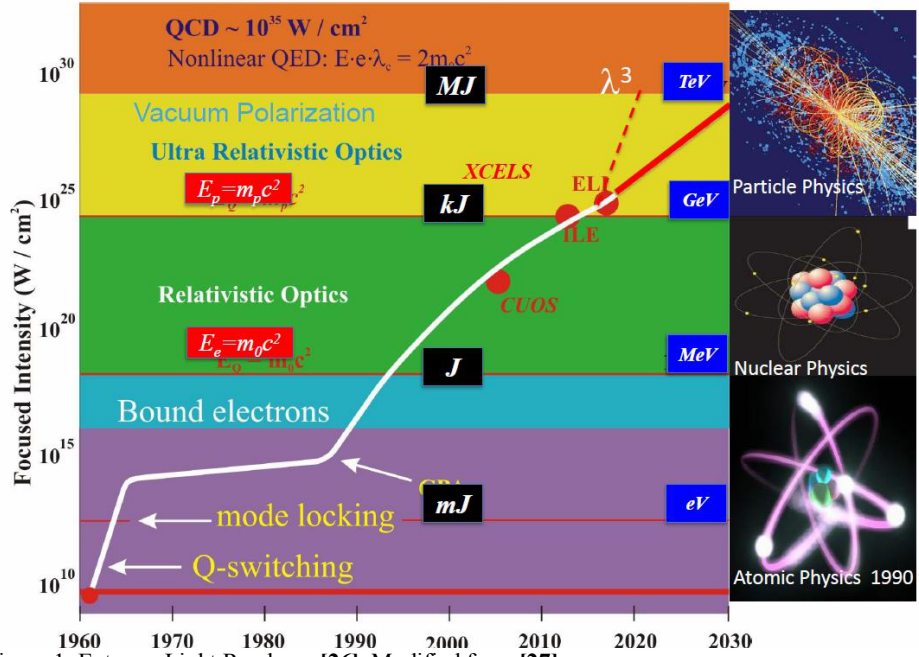

Figure 1: Extreme Light Roadmap [26]. Modified from [27].

An additional factor which can lead to greater-than-expected particle production rates is the exact mathematical form of the driving laser. Rigorous derivations of the exact mathematical form of a monochromatic EM pulse in a vacuum according to a finite power requirement are in the literature [28], [29]. Such an EM pulse has an energy density oscillation at a frequency equal to twice the wave frequency [30]. If the laser is not entirely monochromatic, the effect on the resulting physical processes of particle production are subtle. However, these considerations, while noteworthy, further reduce, rather than enhance, the particle production rate.

Considered here is a hypothetical vaporization laser with an electric field in excess of the Schwinger Limit. Although the first lasers with electric field strengths in excess of the critical value are most likely to be *x*-ray free electron lasers or gamma ray lasers, this scenario considers the future prospect of a coherent beam of 1 μm photons whose intensity is at least the Schwinger Limit. In [31], it is shown that for a temporally localized (pulsed) electric field *E(t)* in the *z*-direction with maximum value $E_0$ and of effective time *T*, such that

$$E_0 T = \int_{-\infty}^{+\infty} E(t) dt, \qquad (7)$$

the critical parameter for particles of mass *m* and charge *q* is $\epsilon_t = \frac{m}{qE_0 T}$. Pair production is strongly suppressed for $\epsilon_t \gg 1$. An applicable example is the Sauter electric field in the *z*-direction defined by

$$E(t) = E_0 [\text{sech}(t/T)]^2, \qquad (8)$$

given by the Sauter potential,

$$A_0(t) = -E_0 T \tanh\left(\frac{t}{T}\right). \tag{9}$$

The particle production density (per unit spatial volume) of charged fermions from the WKB approximation up through quartic terms was derived in [31] as:

$$N^{(0)} = \frac{(qE_0)^{5/2}T}{4\pi^3 m}(1+\epsilon_t^2)^{\frac{5}{4}}e^{-Z_t\left(\sqrt{1+\epsilon_t^2}-1\right)}\left[1 - \frac{5}{16}(4-3\epsilon_t^2)b_t\right], \tag{10}$$

where $Z_t \equiv 2\pi q E_0 T^2 = \frac{2}{\delta \epsilon_t^2}$, $\delta \equiv \frac{qE_0}{\pi m^2}$, and $b_t \equiv \frac{\delta}{\sqrt{1+\epsilon_t^2}}$. As Kim and Page point out, "the WKB instanton action plus the next-to-leading order contribution in spinor QED equals the WKB instanton action in scalar QED, thus justifying why the WKB instanton in scalar QED can work for the pair production of fermions" [31]. This implies that the WKB approximation is valid, and $b_t \ll 1$ serves as an adiabaticity parameter. In this limit, the particle production rate simplifies to,

$$N^{(0)} \approx \frac{(qE_0)^{\frac{5}{2}}T}{4\pi^3 m}(1+\epsilon_t^2)^{\frac{5}{4}}e^{-Z_t\left(\sqrt{1+\epsilon_t^2}-1\right)}. \tag{11}$$

In the presence of a constant magnetic field ***B*** that is parallel to the electric field ***E*** (as in the case of two perpendicular lasers), the particle production rate is modified to

$$N^B = \frac{qBc(qE_0)^{3/2}T}{4\pi^2 m}(1+\epsilon_t^2)^{\frac{3}{4}}e^{-Z_t\left(\sqrt{1+\epsilon_t^2}-1\right)} \coth\left(\frac{\pi cB}{E_0(1+\epsilon^2)^{1/2}}\right). \tag{12}$$

(Note that, since $\lim_{B\to 0} B \coth\left(\frac{\pi cB}{E_0(1+\epsilon_t^2)^{1/2}}\right) \to E_0(1+\epsilon_t^2)^{1/2}$, then $\lim_{B\to 0} N^B \to N^{(0)}$ as expected—i.e., (12) reduces to (11).) In the $\varepsilon \approx 0$ (in the large $E_0 T$) limit, the electron/positron particle production density in count per unit volume approaches,

$$N^B(T) \approx \frac{eBc(eE_0)^{3/2}T}{4\pi^2 m} \coth\left(\frac{\pi cB}{E_0(1+\epsilon_t^2)^{1/2}}\right). \tag{13}$$

(Note that the magnitude of the ***B***-component from a second field is generically independent of the magnitude of ***E***₀ of the first field, unless the two sources are identical, such as two matching lasers. In the latter case, one can replace ***B*** with ***E***₀/*c*.)

Electron-positron pair production rate (per unit area per unit second) corresponds to the pulse time *T* in the particle production density being replaced by the corresponding pulse length scale *L* = *Tc*.

$$N^B(L) \approx \frac{eBc(eE_0)^{3/2}L}{4\pi^2 m} \coth\left(\frac{\pi cB}{E_0(1+\epsilon_t^2)^{1/2}}\right). \tag{14}$$

Likewise, (14) also nearly expresses the electron-positron pair production rate for an electric field *E(x)* spatially localized along the *x*-direction (and pointing along the *z*-direction) with maximum value *E₀*, and of effective length *L* (this is the laser beam waist) defined by

$$E_0 L = \int_{-\infty}^{+\infty} E(x)dx, \tag{15}$$

for the Sauter electric field

$$E(x) = E_0[\text{sech}(x/L)]^2. \tag{16}$$

The only modification then required of (14) is that $\epsilon_t^2$ be replaced by $-\epsilon_L^2$, where $\epsilon_L = \frac{m}{qE_0L}$.

From eq. (13) (or equivalently eq. (14)), the pair production fluxes emerging from the laser spot are calculated for a futuristic 1.0 attosecond ($10^{-18}$ s) 1 μm pulse laser producing a spot size of 1μm and with electric field strengths in excess of the Schwinger Limit, and are shown in Figure 2. While the choice of the cross-section being equal to the diffraction limit (proportional to the laser wavelength) is logical, such a spot size is inappropriate for cutting/vaporization lasers. Even a spot size of 1 μm is extremely optimistic, as keyhole behavior between the "Rosenthal"[3] and "Humping"[4] regimes is not well-established for spot sizes much smaller than several hundred microns. However, a keyhole of insufficient diameter is likely to be flooded by chaotic fluctuations and liquid swellings of the surrounding molten material.

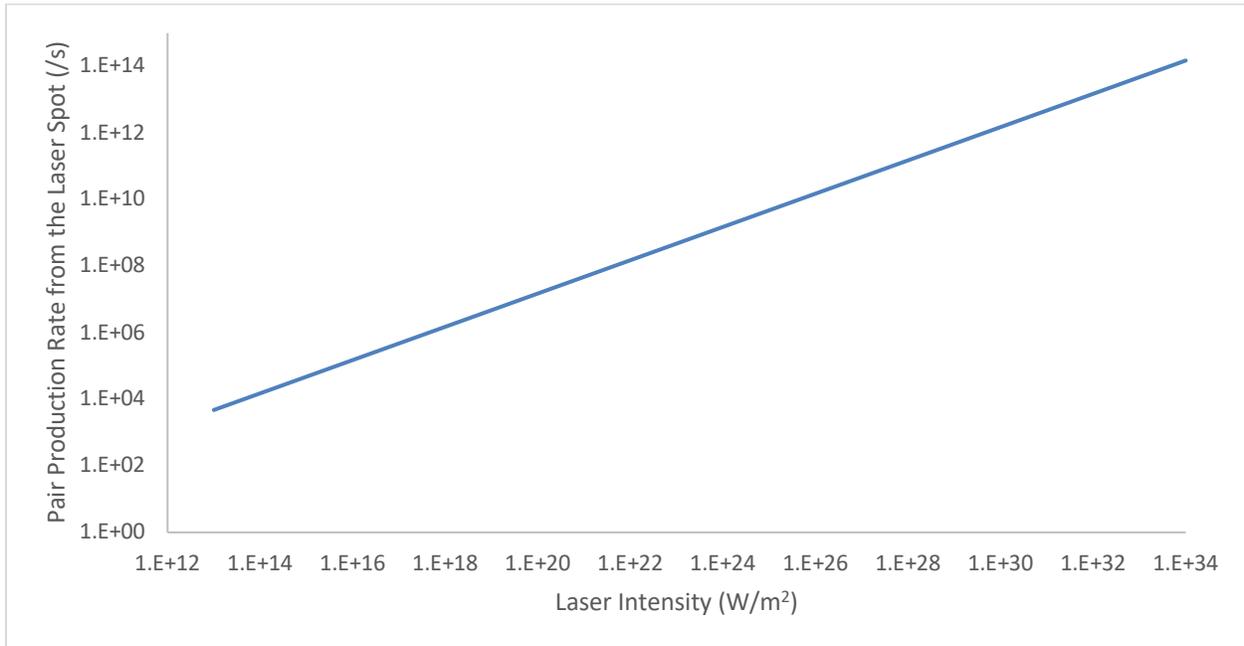

Figure 2: Pair production rate from the laser spot as a function of laser intensity for a 1 attosecond pulse laser with a 1 μm spot size.

The results from Figure 2 are several orders of magnitude below the particle production rates in Table 1, indicating that the vast majority of the particle pairs in the Astra Gemini and Vulcan PW data are produced from γ-γ interactions, and do not result from an optical analog to Hawking radiation. To determine whether the particles bled from the quantum vacuum for laser pulses above the Schwinger Limit constitute a significant energy depletion, the rate at which energy is lost to particle production must be determined.

---

[3] The welding regime in which material welding speeds are slower than ~5 m/min. The large melt pool that forms at the front of the keyhole experiences numerous chaotic surface fluctuations and large liquid swellings in the vicinity of the keyhole aperture.

[4] The welding regime in which material welding speeds exceed ~20 m/min and for which exceptionally strong undercuts, made up of large ellipsoid swellings and separated by comparatively small valleys, occur along the weld seam.

The corresponding laser power to Figure 2 is shown in Figure 3.

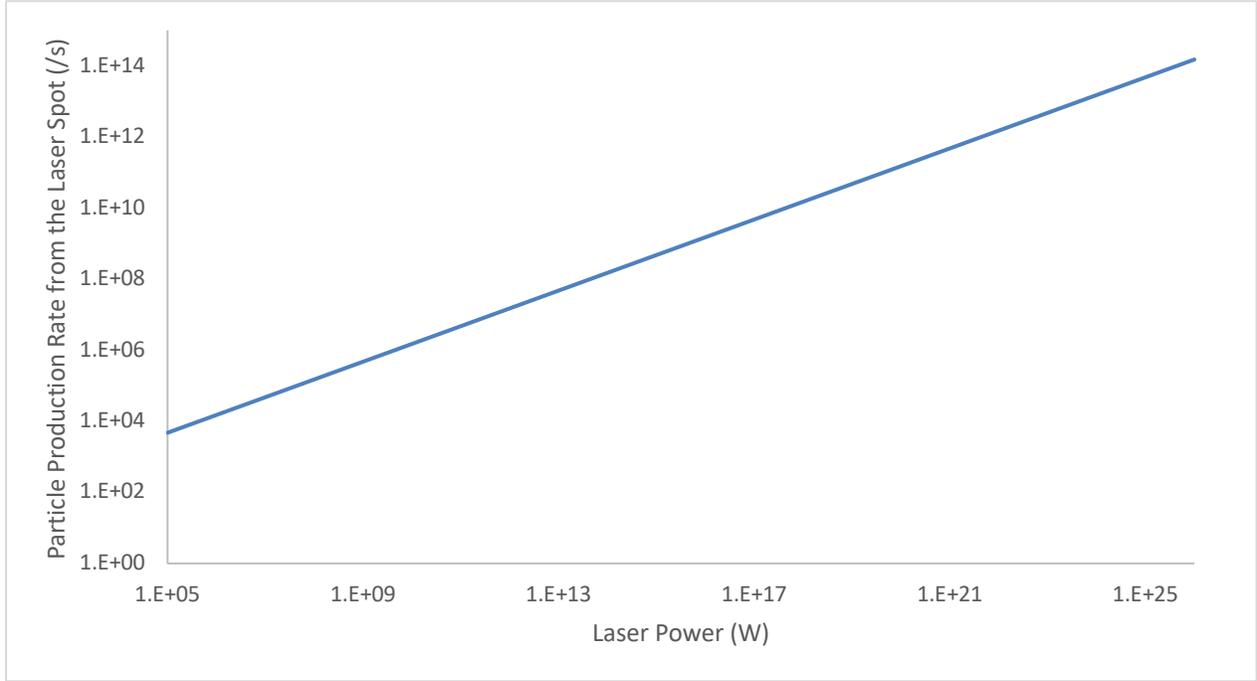

Figure 3: Particle production rate from the laser spot versus laser power for a 1 attosecond pulse laser with a 1 μm spot size.

The power absorption of individual photons by the quantum vacuum to produce particle pairs is $2h\nu N$. This yields a highly negligible power loss of $4.6 \times 10^{-11}$ W, corresponding to a 1 μm laser of $10^{26}$ W. For a hypothetical 1 fm, $10^{26}$ W gamma ray laser (corresponding to a photon energy of 100 GeV), also with a 1 μm spot size, the power loss due to bleeding the quantum vacuum is 370 mW. If the spot size could be reduced to 1 fm, the power realized from the quantum vacuum would be $3.70 \times 10^{17}$ W, which is still vastly below the laser power. In each case, the spot size and laser wavelength are both clearly below the lower limits for keyhole formation by conductive processes in laser materials processing applications.

The results and data presented here are based on Kim's and Page's approach to standard relativistic quantum field theory [31]. Even if a slight improvement in the physical picture is desired by means of an improvement in the many-body technique, it will be future task, and this standard quantum field theory is basically reliable to deal with pair-production.

Additionally, it is not anticipated that external enhancement of the magnetic field would significantly increase the pair production rate because $N$ scales linearly with **B**. Even the magnetic field of a magnetar, which is expected to be $(3 \text{ to } 4) \times 10^{15}$ Gauss [32], would increase the particle production rate by 11 orders of magnitude, which would elevate the power bled from the quantum vacuum to two orders of magnitude in excess of the laser power.

## 6. Summary

The quantum vacuum has been shown to be only an extremely negligible sink of vaporization laser energy for both subcritical and supercritical intensities. Even well beyond the Schwinger Limit, where non-linear QED and non-linear QCD are dominant processes, lasers with spot sizes, beam wavelengths, and pulse durations which are significantly too short (or too small) for conductive keyhole formation, liberate a total power from the quantum vacuum that is substantially smaller than the laser power. As a result of the approximately linear dependence of the pair production rate on the magnetic field, external *B*-field augmentation results in only a negligible increase in the pair production rate.

## 7. Acknowledgements